# Suppression of Coriolis error in weak equivalence principle test using $^{85}$Rb-$^{87}$Rb dual-species atom interferometer[*]

Wei-Tao Duan (段维涛)[1,2], Chuan He (何川)[1,2], Si-Tong Yan(闫思彤) [1,3], Yu-Hang Ji(冀宇航)[1,2], Lin Zhou (周林) [1,2,3†], Xi Chen (陈曦)[1,2,3], Jin Wang (王谨)[1,2,3‡], and Ming-Sheng Zhan (詹明生)[1,2,3]

[1]*State Key Laboratory of Magnetic Resonance and Atomic and Molecular Physics, Innovation Academy for Precision Measurement Science and Technology, Chinese Academy of Sciences, Wuhan 430071, China*

[2]*School of Physics, University of Chinese Academy of Sciences, Beijing, 100049, China*

[3]*Center for Cold Atoms Physics, Chinese Academy of Sciences, Wuhan 430071, China*

Coriolis effect is an important error source in the weak equivalence principle (WEP) test using atom interferometer. In this paper, the problem of Coriolis error in WEP test is studied theoretically and experimentally. In theoretical simulation, Coriolis effect is analyzed by establishing an error model. The measurement errors of Eötvös coefficient ($\eta$) in WEP test related to experimental parameters, such as horizontal-velocity difference and horizontal-position difference of atomic clouds, horizontal-position difference of detectors and rotation compensation of Raman laser's mirror are calculated. In experimental investigation, the position difference between $^{85}$Rb and $^{87}$Rb atomic clouds is reduced to 0.1 mm by optimizing the experimental parameters, an alternating detection method is used to suppress the error caused by detection position difference, thus the Coriolis error related to atomic clouds and detectors is eliminated to $1.1\times10^{-9}$. This Coriolis error is further corrected by compensating the rotation of Raman laser's mirror, and the total uncertainty of $\eta$ measurement related to Coriolis effect is reduced as $\delta\eta = 4.4 \times 10^{-11}$.

## 1. Introduction

Atom interferometer (AI) [1] is a new type of precision measuring instrument, which has been widely used in precision measurement of gravity, [2-11] gravity gradient, [8, 12-14] rotation [15-19] and test of weak equivalent principle (WEP).[20-29] Coriolis effect is an important error source in gravity measurement experiments using AIs. [11, 30-34] To compensate the effect of the Coriolis effect in atom interferometry, several effective methods have been used in single-species AI-based gravity measurements. Peters *et al*. [33] used the interference signal itself to find the correct position of the detection region (where the Coriolis effect vanishes) by introducing additional, well-defined rotations. Louchet-Chauvet *et al*. [34] corrected Coriolis shift by carrying out measurements with opposite orientations with respect to the Earth's rotation vector direction. Lan *et al*.[35] used a piezoelectric tube (PZT)-controlled tip-tilt mirror to regulate the orientation of the Bragg beam and compensate for the Earth's rotation. Different from the configuration of the single-species AI, the dual-species AI used for WEP tests can achieve time- or space-synchronized interference measurement using different species atoms, which has good performance in common-mode-noise suppression. However, due to the differences in the initial positions and velocities of the dual-species atomic clouds, the Coriolis effect related to the horizontal motion of the atomic clouds is still a major error source in AI-based WEP tests.[21, 22] In $^{85}$Rb-$^{87}$Rb dual-species AI, we achieved common-mode-noise suppression by the four-wave double-diffraction Raman transition (4WDR) scheme,[24] which greatly reduced contributions from laser phase noise, ac Stark shift and quadratic Zeeman shift. But, the Coriolis effect error, which was introduced by the small inconsistent of the centroid positions and velocities of $^{85}$Rb and $^{87}$Rb atomic clouds, didn't be effectively suppressed, the uncertainty of this error is as high as $2.9\times10^{-8}$,[24] which was the largest error term in the experiment at that time. To effectively improve the precision of WEP test using the 4WDR dual-species AI, we study the suppression of Coriolis error in this paper. We first theoretically analyze the phase shift factors related to the Coriolis effect in the 4WDR AI, then calculate the Coriolis errors related to experimental parameters, such as the velocity differences, position differences of atomic clouds, the position differences of detectors, and the rotation rate of Raman laser's mirror. Based on the analysis, we find experimental means to suppress the Coriolis effect. By optimizing the background magnetic field, the parameters of cooling laser beams and the loading time of magneto-optical trap (MOT), the position difference between $^{85}$Rb and $^{87}$Rb atomic clouds is reduced. By using the alternating detection method (ADM), the uncertainty caused by the position fluctuation of dual-species atomic clouds is suppressed. By two-dimensional (2D) rotation compensation for Raman laser's mirror, the total uncertainty of $\eta$ measurement related to Coriolis effect is reduced to $\delta\eta=4.4\times10^{-11}$.

[*] Project supported by National Key Research and Development Program of China under Grant No. 2016YFA0302002, National Natural Science Foundation of China under Grant Nos. 91736311, 11574354, Strategic Priority Research Program of the Chinese Academy of Sciences under grant No. XDB21010100 and the Youth Innovation Promotion Association of the Chinese Academy of Sciences under grant No. 2016300.
† Corresponding author. E-mail: lzhou@wipm.ac.cn
‡ Corresponding author. E-mail: wangjin@wipm.ac.cn

## 2. Theoretical model

To quantitatively investigate the Coriolis error in the 4WDR AI and provide guidance for experiments, we establish an error model to analyze the dependence of Coriolis phase shift on the velocity, position of atomic clouds and the position of detectors(the velocity and position discussed in this paper are in the horizontal direction), then theoretically simulate the performance of rotation compensation of Raman laser's mirror.

### 2.1. Error model

The Eötvös coefficient in $^{85}$Rb-$^{87}$Rb dual-species AI-based WEP test is [24]

$$\eta=\frac{2(g_{87}-g_{85})}{g_{85}+g_{87}}, \tag{1}$$

where $g_i$ ($i$=87, 85) is the gravitational acceleration of $^{87}$Rb or $^{85}$Rb atoms. The phase shift is simply described as

$$\phi_i = k_{\text{eff}}^i g_i T^2, \tag{2}$$

where $k_{\text{eff}}^i$ ($i$=87, 85) is the effective wave vector of Raman laser for $^{87}$Rb or $^{85}$Rb atoms,

and $T$ the interval between Raman pulses. According to Eqs. (1) and (2), the relation between $\eta$ and phase shifts is

$$\eta = \frac{2\left(\phi_{87} k_{\text{eff}}^{85} - \phi_{85} k_{\text{eff}}^{87}\right)}{\phi_{87} k_{\text{eff}}^{85} + \phi_{85} k_{\text{eff}}^{87}}. \quad (3)$$

Usually, the value of $\eta$ measured by experiments consists of three parts, $\eta = \eta_0 + \Delta\eta \pm \delta\eta$, where $\eta_0$ is the ideal value of the Eötvös coefficient, $\Delta\eta$ the systematic error and $\delta\eta$ the uncertainty of $\Delta\eta$.

The schematic diagram of $^{85}$Rb-$^{87}$Rb dual-species AI is shown in Fig. 1, where Fig.1 (a) is the schematic diagram of relevant hyperfine-levels of rubidium atoms. Take the $^{85}$Rb atom as an example to briefly explain the atom interference process. As shown in Fig. 1(b), the atoms in initial state $|F=2\rangle$ interact with $\pi/2$-$\pi$-$\pi/2$ Raman pulses to implement interference loop. The wave vectors of three Raman lasers are $k_1$, $k_2$ and $k_3$. The atom interference loop is spatially symmetric because Raman pairs ($k_1$, $k_3$) and ($k_2$, $k_3$) supply recoil momentum in opposite directions. The first $\pi/2$-pulse is for transferring atoms from state $^{85}$Rb$|F=2, 0\rangle$ to state $^{85}$Rb$|F=3, \hbar k_{\text{eff}}\rangle$ and $^{85}$Rb$|F=3, -\hbar k_{\text{eff}}\rangle$ at site A, where $\hbar k_{\text{eff}}$ and $-\hbar k_{\text{eff}}$ are recoil momentum of atoms obtained from Raman pulse. The $\pi$-pulse is for reflecting atoms in different monument states $^{85}$Rb$|F=3, \hbar k_{\text{eff}}\rangle$ and $^{85}$Rb$|F=3, -\hbar k_{\text{eff}}\rangle$ at sites B and C, where the population of different monument states exchange. The second $\pi/2$-pulse is for combing atoms in states $^{85}$Rb$|F=3, \hbar k_{\text{eff}}\rangle$ and $^{85}$Rb$|F=3, -\hbar k_{\text{eff}}\rangle$ at site D. The detected population of $^{85}$Rb$|F=2\rangle$ or $^{85}$Rb$|F=3\rangle$ at D depends on the phase shifts of Raman pulses. Similarly, for $^{87}$Rb atoms, the initial state is $^{87}$Rb$|F=1, 0\rangle$ and the interference states are $^{87}$Rb$|F=2, \hbar k_{\text{eff}}\rangle$ and $^{87}$Rb$|F=2, -\hbar k_{\text{eff}}\rangle$. The wave vectors of three Raman lasers are $k_1$, $k_2$ and $k_4$.

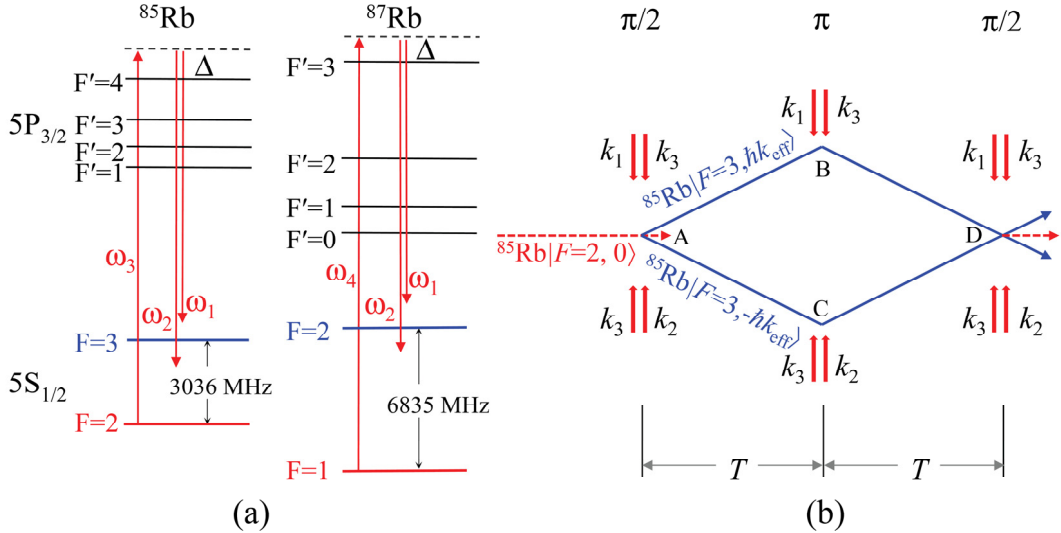

**Fig.1** (color online) Schematic diagram $^{85}$Rb-$^{87}$Rb dual-species atom interferometer. (a) Related sub-levels of rubidium atoms. Raman lasers with frequencies of $\omega_1$, $\omega_2$ and $\omega_3$ are used for $^{85}$Rb atoms, while that with frequencies $\omega_1$, $\omega_2$ and $\omega_4$ are for $^{87}$Rb atoms; $\omega_1$ and $\omega_2$ are detuned to the blue side of transitions $^{85}$Rb$|F = 3\rangle$ to $|F'= 4\rangle$ and $^{87}$Rb$|F =2\rangle$ to $|F'= 3\rangle$. (b) Schematic diagram of 4WDR interference path for $^{85}$Rb atoms, $k_1$, $k_2$ and $k_3$ are wave vectors of Raman beams $\omega_1$, $\omega_2$ and $\omega_3$, respectively. For $^{87}$Rb atoms, states $^{87}$Rb$|F=1\rangle$ and

$^{87}$Rb|F=2⟩ are involved, and $k_3$ is replaced by $k_4$.

In the local laboratory (Wuhan, 30.54°N) coordinate system, assume that the positive direction of the z-axis is upward along the plumb line. The y-axis is the direction of the unit vector in the horizontal plane from south to north, x-axis is the direction of the unit vector in the horizontal plane from west to east. Considering the Earth as a regular sphere, the rotation angular vector of the Earth is written as

$$\vec{\Omega}_E = (0, \Omega_y, \Omega_z). \tag{4}$$

The projection of the Earth's rotation rate in the laboratory's y-axis is $\Omega_y$. We define $\Omega_y = \Omega_0$ as the reference point for rotation compensation, $\Omega_0 = 62.8$ μrad/s. The trajectories of atoms are expanded as time-varying polynomials

$$\vec{r}(t) = \sum \vec{r}_i (t-t_0)^i, \tag{5}$$

where, $\vec{r}_i$ (i=0, 1, 2, …) are time-independent parameters. The dynamic equation of the atom is [36]

$$\frac{\partial^2 \vec{r}}{\partial t^2} = \vec{g} - 2\vec{\Omega}_E \times \frac{\partial \vec{r}}{\partial t} - \vec{\Omega}_E \times \left[\vec{\Omega}_E \times (\vec{r} + \vec{R})\right] + \Gamma \cdot \vec{r}(t), \tag{6}$$

where $\vec{g}$ is the local gravitational acceleration, $\vec{R}$ the connecting line between the Earth's center point and the coordinate origin of the laboratory system, and $\Gamma$ the local gravity gradient tensor. The total Coriolis phase shift of the interferometer is

$$\phi = -2\vec{k}_{eff} \cdot (\vec{\Omega}_E \times \vec{v}_0) T^2, \tag{7}$$

where $\vec{v}_0$ is the average horizontal-velocity of the atoms.

In the experiment, the average velocity of atomic clouds measured by a detector is related to the detection position. [33, 37] To quantitatively describe the velocity distribution of atoms in the detection area, we directly calculate the probability of atoms in the detection area. At the initial moment, the state of atom can be described by the product of the initial velocity distribution and the initial position distribution. Assume the atoms are inside the detection area at time $t$, the convolution (that is the equivalent velocity distribution of the atoms in the detection area) of atom's position and velocity can be obtained by full-space integration. Thus, we calculate the Coriolis phase shift in differential-measurement of gravity using dual-species AI under various atomic parameters and detection parameters.

The Coriolis effect was compensated by changing the orientation of Raman laser's wave-vector in single-species AI. [35] Here, we apply this rotation compensation method to the dual-species AI. Considering that the incident Raman beam is reflected by a PZT-driven mirror. Let $\vec{k}_i$ be the wave-vector of incident beam, $\vec{k}_o$ the wave-vector of exit beam and $\vec{n}$ the unit normal vector of the mirror, then $\vec{k}_o = \vec{k}_i - 2(\vec{k}_i \cdot \vec{n})\vec{n}$.[38] In the mirror coordinate system, the west-east diameter in the mirror surface is defined as x' axis, and the diameter orthogonal to both x'-axis and $\vec{n}$ is y' axis. When the 2D compensation angle $(\theta_{x'}, \theta_{y'})$ is applied to each Raman pulse, the compensated phase shift is

$$\phi = \vec{k}_o(0,0) \cdot \vec{r}(0) - 2\vec{k}_o(\theta_{x'}, \theta_{y'}) \cdot \vec{r}(T) + \vec{k}_o(2\theta_{x'}, 2\theta_{y'}) \cdot \vec{r}(2T), \tag{8}$$

where $\vec{r}(0)$, $\vec{r}(T)$ and $\vec{r}(2T)$ are atomic positions at time 0, T and 2T, respectively.

$\theta_{x'}$ and $\theta_{y'}$ are rotation angles of the PZT around x′ and y′ axis, respectively.

In the laboratory coordinate system, the actual rotation compensation performance can be evaluated by the contrast of interference fringes, [35] that is, when the fringe contrast reaches the maximum value, the corresponding mirror rotation rate just compensates the Earth's rotation effect, the relevant phase shift is represented as

$$\phi = \phi_o + c_1 v_x + c_2 v_y + c_3 x + c_4 y, \qquad (9)$$

where $c_i$ ($i$=1, 2, 3, 4) is the transformation coefficient of parameter $i$, $v_x$, $v_y$ are velocities of atoms in the x and y directions, respectively. $x$ and $y$ are positions of atoms in x and y directions, respectively. The fringe contrast is expressed as

$$C = \exp[-(c_1^2 \sigma_{v_x}^2 + c_2^2 \sigma_{v_y}^2 + c_3^2 \sigma_{x_0}^2 + c_4^2 \sigma_{y_0}^2)/2], \qquad (10)$$

where $\sigma_j$ ($j=v_x$, $v_y$, $x_0$, $y_0$) is the standard deviation of Gaussian distribution of $j$. In the following, we consider the actual experimental conditions and analyze the Coriolis error in $^{85}$Rb-$^{87}$Rb dual-species AI-based WEP test.

### 2.2. Parameters of atomic clouds and Coriolis error

Due to the differences in mass and hyperfine levels, the average velocity, center position, size and temperature of $^{85}$Rb and $^{87}$Rb atomic clouds are also different. Suppose the atoms are launched 2 m up and fall to the detection area after 1.2 s, the area of the detection is 10 mm × 10 mm, two detectors for $^{85}$Rb and $^{87}$Rb atoms are 30 mm apart, and the atomic clouds are symmetrically distributed in the vertical direction. We calculate the dependence of Coriolis shift on the difference of initial velocity and position between $^{85}$Rb and $^{87}$Rb atomic clouds.

The simulated error of $\eta$ versus atomic cloud parameters are shown in Fig. 2. Fig. 2 (a) shows the dependence of $\Delta\eta$ on the velocity difference between $^{85}$Rb and $^{87}$Rb atomic clouds, where, the temperature of atomic clouds are selected as 10 nK, 100 nK and 3 μK, the size of atomic clouds are selected as 0.1 mm, 1 mm, 3 mm and 5 mm. Fig. 2 (b) shows the dependence of $\Delta\eta$ on the position difference between $^{85}$Rb and $^{87}$Rb atomic clouds under various size and temperature values. It can be seen that with the increase of the temperature and size of atomic clouds, $\Delta\eta$ varies differently with the initial velocity difference and position difference. For ultra-cold (10 nK, 0.1 mm) $^{85}$Rb and $^{87}$Rb atomic clouds, the Coriolis effect caused by initial velocity difference is much larger than that of cold (3 μK, 3 mm) atomic clouds, while the Coriolis effect caused by position difference is smaller than that of cold (3 μK, 3 mm) atomic clouds. This is because when atoms freely diffuse for 1.2 s, the sizes of 10 nK clouds are small, and the average velocity difference between two atomic clouds at detection area does not change much from that at launching point, while the 3 μK clouds have large size which are selectively detected by position-fixed detectors, only $^{85}$Rb and $^{87}$Rb atoms with smaller velocity difference synchronously enter the detection area. Therefore, the $\Delta\eta$ value simulated by 3 μK $^{85}$Rb and $^{87}$Rb atoms varies slowly with the initial velocity difference. However, when the initial positions are different, the average horizontal-velocities of 3 μK $^{85}$Rb and $^{87}$Rb atomic clouds which entering the same detection area are different, and their difference is much larger than that of 10 nK clouds, the error of $\eta$ value simulated by 3 μK $^{85}$Rb and $^{87}$Rb atoms varies quickly with the initial position difference. For typical $^{85}$Rb and $^{87}$Rb atomic clouds with size of 3 mm, temperature of 3 μK, position difference of 0.1 mm and velocity difference of 1 mm/s, the systematic error of Coriolis effect is $1.1\times10^{-9}$ and that caused by velocity difference is $-5.8\times10^{-10}$.

In practical experiment, when the size of the detection area is fixed, for ultra-cold $^{85}$Rb and $^{87}$Rb atomic clouds with an initial temperature of 10 nK, it is necessary to accurately control and reduce their velocity difference to suppress Coriolis error. For cold atomic clouds with an initial temperature of several microkelvin, it is necessary to precisely control and reduce their position difference to eliminate Coriolis error.

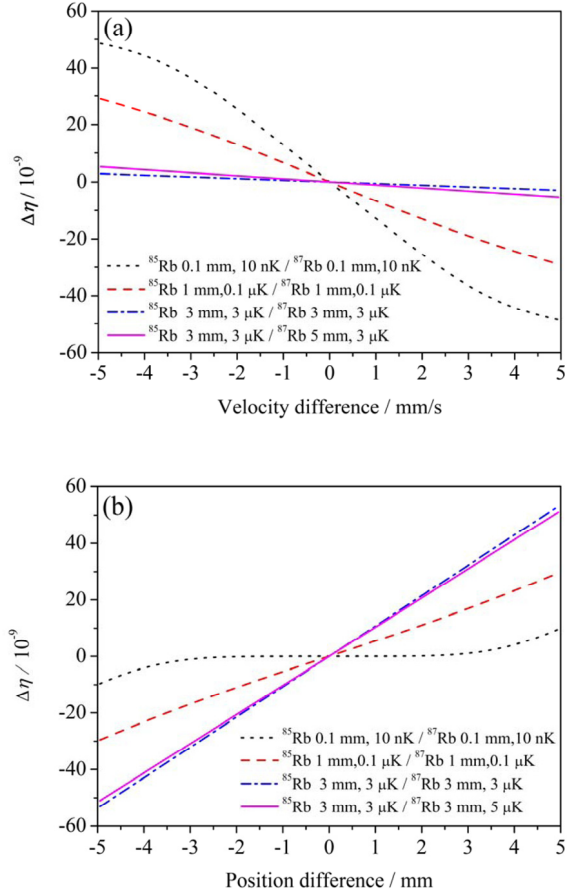

**Fig. 2** (color online) Dependence of $\Delta\eta$ on initial parameters of dual-species atomic clouds. (a) $\Delta\eta$ versus initial velocity difference between $^{85}$Rb and $^{87}$Rb atomic clouds. (b) $\Delta\eta$ versus initial position difference between $^{85}$Rb and $^{87}$Rb atomic clouds. The dot line is the data for $^{85}$Rb and $^{87}$Rb atomic clouds with size of 0.1 mm and temperature of 10 nK, the dash line is for 1 mm, 0.1 μK atomic clouds, the dash-dot line is for 3 mm, 3 μK atomic clouds, the solid line is for 3 mm, 3 μK $^{85}$Rb atomic cloud and 5 mm, 3 μK $^{87}$Rb atomic cloud.

## 2.3. Detector location and Coriolis error

Since the Coriolis effect is proportional to the horizontal-velocity of atomic clouds, different detecting position leads to a difference in horizontal-velocity selection. This detection effect [37] introduces Coriolis error. Assuming $^{85}$Rb and $^{87}$Rb atomic clouds are detected simultaneously by two different detectors, we simulate the dependence of $\Delta\eta$ on the position difference of detection, the results are shown in Fig. 3. When the initial position and velocity of the atomic clouds are determined (corresponding to the case where the temperature is 3 µK and the size of the detection area is 10 mm ×10 mm), if the detection positions of $^{85}$Rb and $^{87}$Rb atoms differ by 1 mm, then the measurement error of $\eta$ value is about 2 ×10$^{-9}$. If we exchange two detectors for $^{85}$Rb and $^{87}$Rb atoms and average the data collected before and after exchanging, the average value is no longer sensitive to the detection position, the measurement error of $\eta$ value is ignorable as shown by solid line in Fig. 3. This means, under the ideal condition where the initial positions of the atomic clouds completely overlap, the ADM can effectively suppress the error caused by the detection position.

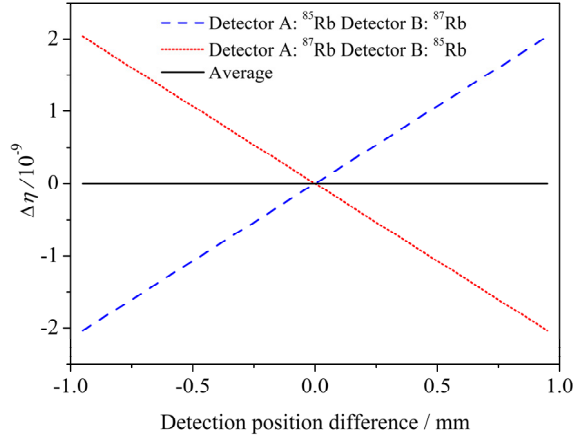

**Fig. 3** (color online) Error of $\eta$ versus the difference of detection position. The dash and dot lines are case before and after alternating detectors, the solid line is the average data for both.

## 2.4. Suppressing Coriolis error by rotation compensation

To effectively correct the Coriolis errors introduced by the position fluctuation of atomic clouds, the initial position difference and initial velocity difference of dual-species atomic clouds, it is necessary to compensate the tilt angle of Raman laser's mirror caused by the Earth's rotation. Let $\Omega_m$ represent the actual rotation rate of Raman laser's wave vector around y-axis, the theoretically simulated relation between $\Omega_m$ and the fringe contrast is shown in Fig. 4, where all atoms are assumed to participate in the interference process and contribute to the signal. It shows that the rotation compensation of Raman laser's mirror corrects the total Coriolis error and significantly improves the fringe contrast. When $\Omega_m$ = -$\Omega_0$, the fringe contrasts of atoms with different temperature reach the maximum value. When $\Omega_m \neq$ -$\Omega_0$, the fringe contrast decreases rapidly with increasing of atom temperature. For atoms with temperature of 3 µK, if no rotation compensation is performed ($\Omega_m$ = 0), the fringe contrast is close to 0. For atoms with temperature of 0.3 µK, the fringe contrast is still more than 60% even without rotation compensation.

The simulated dependence of $\Delta\eta$ on mirror's rotation rate is shown in Fig. 5, where both the position difference of atomic clouds and the inconsistencies of detectors are considered, the temperature of atoms is 3 µK, the size of atomic clouds is 3 mm, and their horizontal center position deviates from 0.1 mm. According to Fig. 5, when $\Omega_m$ = -$\Omega_0$, the curves obtained by two detection methods intersect, and the ordinate of the crossing point is 0. That is, at the ideal value of rotation compensation, the fringe contrast is the best, the error of $\eta$ measurement is the smallest. From Fig. 5 we find that the errors caused by both detection methods (dot line and dash lines) and the mismatch of atomic clouds (slid line) have linear relationship with the mirror rotation angle, therefore, the optimal rotation rate can be determined by the fringe contrasts.

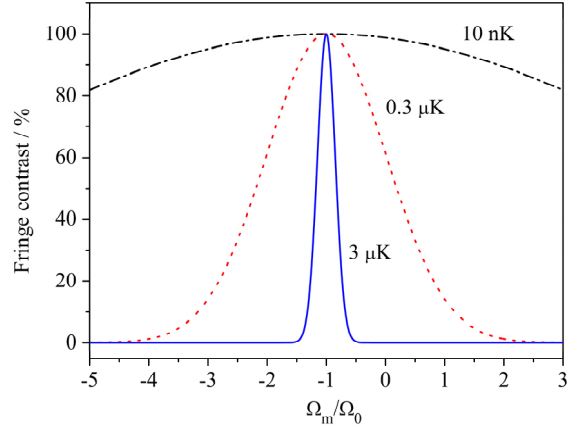

**Fig. 4** (color online) Fringe contrast ($T$=203 ms) versus mirror's rotation rate. The influence of the detector was not considered here. The dash-dot line is for atoms with temperature of 10 nK, the rotation compensation is not obvious. The dot line is for 0.3 μK atoms, the rotation compensation is effective in the range of -4 $\Omega_0$ ~ + 2 $\Omega_0$, the solid line is for 3 μK atoms, the rotation compensation is effective in the range of -1.5 $\Omega_o$ ~ + 0.5 $\Omega_0$.

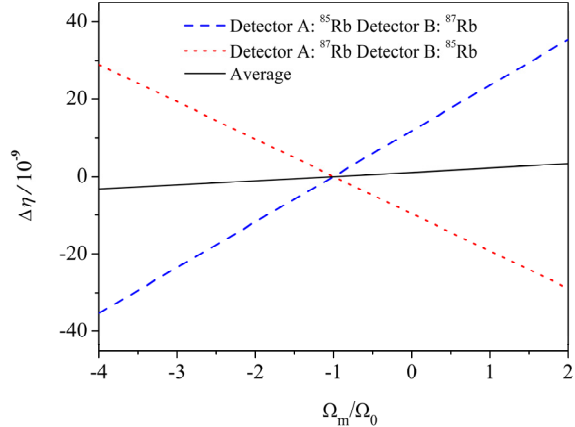

**Fig. 5** (color online) Error of $\eta$-value versus rotation rate of the Raman laser's mirror. Dash and dot lines are data for alternating detectors. Solid line is the average data of two detection methods, which is the contribution of position difference of $^{85}$Rb and $^{87}$Rb atomic clouds.

We also simulate the rotation compensation around y-axis for Raman laser's mirror. The dependence of fringe contrast on the interval of Raman pulses ($T$) is shown in Fig.6. If $T$ = 200 ms, the fringe contrast with rotation compensation is increased from 67% to 99%; when $T$ = 1 s, the fringe contrast with rotation compensation is increased from 0 to 71%. The rotation compensation is necessary for situations with longer Raman pulse intervals in long baseline AI.

## 3. Experimental measurement

The above simulations have three implications: (1) controlling and reducing the initial position and initial velocity difference of atomic clouds is a necessary means to reduce the Coriolis error; (2) the ADM is useful to reduce the measurement error introduced by detection effect; (3) rotation compensation of Raman laser's mirror is necessary for correcting Coriolis error. The experimental setup and experimental results are described in detail as below.

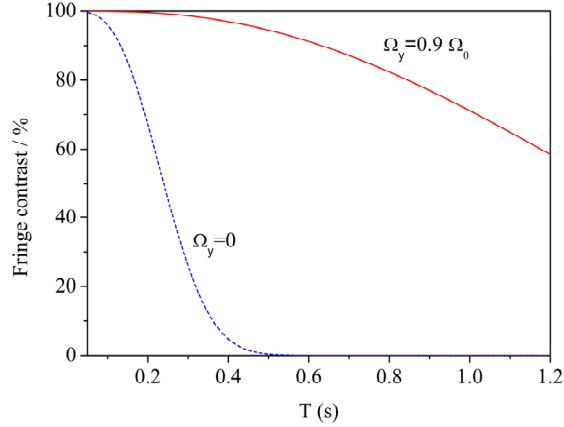

**Fig. 6** (color online) Fringe contrast versus the interval of Raman pulses ($T$). The dash line is the simulation without rotation compensation, and the solid line is the simulation with rotation compensation rate at 0.9 $\Omega_0$.

### 3.1. Experimental setup

In our previous experiments, [24, 39] due to the difference between the initial velocities of $^{85}$Rb and $^{87}$Rb atomic clouds, the resulting Coriolis error was $\Delta\eta = 2.9\times10^{-8}$.[24] To improve the positional coincidence of atomic clouds, reduce detection error, and effectively compensate the Coriolis effect, here we add a 2D-PZT turntable (PI S-330.2SH) to control Raman laser's mirror. The experimental setup is shown in Fig. 7. The Raman laser's mirror is mounted on the 2D-PZT turntable with a cross angle of 45° to the horizontal plane. The 2D-PZT turntable can perform rotation modulation on x′- and y′-axis, it has a rotation resolution of 20 nrad and an angle tuning range of 20 mrad. Four Raman laser beams (with frequencies of $\omega_1$, $\omega_2$, $\omega_3$ and $\omega_4$) are divided into two groups. One group ($\omega_2$, $\omega_3$ and $\omega_4$) is reflected by the mirror and propagate upward from the bottom window of the MOT chamber. Another group ($\omega_1$, $\omega_3$ and $\omega_4$) propagates downward from the top window of the vacuum chamber. The center of the detection area is located 25.4 cm above the center of MOT chamber. Detector A and B are installed on the north and south sides of the detection area, respectively. They are 30 mm apart in the vertical direction.

The MOT simultaneously traps and launches up $^{85}$Rb and $^{87}$Rb atoms to form dual-species atom fountains. The height of the fountains is 2 m. During the launching process, the atoms undergo velocity-selection and state-preparation, they then enter the magnetically shielded region, where the 4WDR scheme is used to implement a dual-species AI. After the interference process, $^{85}$Rb and $^{87}$Rb atoms fall into the detection area and are detected by two detectors. The duration of the first Raman pulse is 31 μs, the interval of Raman pulses is adjusted to 203.164 ms, so that the offset phase difference between the dual-species interference fringes is 20.5 π, this is helpful for fitting the actual phase difference between $^{85}$Rb and $^{87}$Rb atoms by ellipse fitting method. [40]

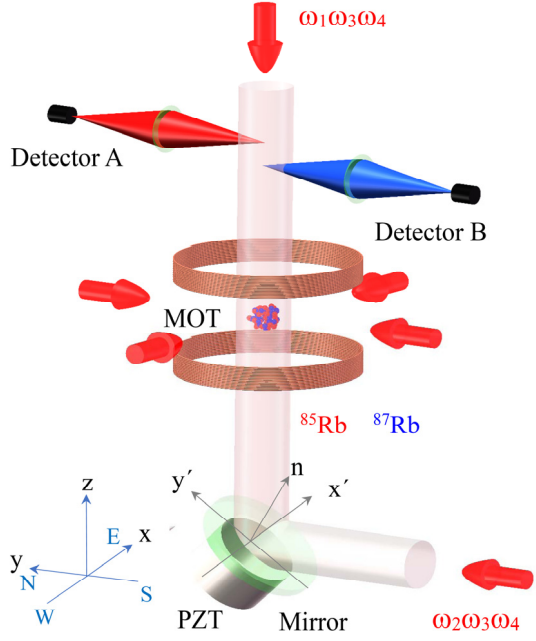

**Fig. 7** (color online) Experimental setup. The Raman laser beams for $^{87}$Rb atoms ($\omega_1$, $\omega_3$ and $\omega_4$) propagate from the top window of vacuum chamber to the bottom and those for $^{85}$Rb atoms ($\omega_2$, $\omega_3$ and $\omega_4$) propagate from the bottom window of vacuum chamber to the top. PZT: piezoelectric tube. MOT: magneto-optical trap.

To improve the coincidence of interference trajectories of the two species atoms, the background magnetic field, the frequency stability and the intensity stability of lasers are optimized. This makes the atomic cloud in the MOT closer to a sphere. In the experiment, the size of $^{87}$Rb atomic cloud increases with MOT loading time, $t$. When $t$ = 2 s, the sizes of $^{87}$Rb and $^{85}$Rb atomic clouds are similar. When $t > 5$ s, the size of $^{87}$Rb atomic cloud is twice that of $^{85}$Rb. Therefore, the loading time of the atoms in the MOT is adjusted to 2 s, so that the position difference of two species of atomic clouds is reduced to the order of 0.1 mm. Atoms undergo vertical-velocity-selection before they enter the interference area, the control accuracy of vertical-velocity is 0.2 mm/s. These means effectively reduce the influence of Coriolis effect, wave-front distortion, gravity gradient and other factors.

In the experiment, the detection error comes from several factors, such as the size and position of detection beam, the angle of fluorescence collection and the installation accuracy of detectors. The installation accuracy of two detectors in west-east direction is better than 1 mm, and that in south-north direction is greater than 1 mm. The angle of the detectors in south-north direction is evaluated and adjusted according to the fluorescence signal. Affected by fluctuation of the magnetic field, laser frequency and laser intensity, the positions of atomic clouds fluctuate. To solve these problems, a real-time ADM is used. In the first detection cycle, Detector A collects the fluorescence signal of $^{85}$Rb atoms, Detector B collects the signal of $^{87}$Rb atoms; in the second detection cycle, two detection beams are exchanged, and Detector A collects the signal of $^{87}$Rb atoms, Detector B collects that of $^{85}$Rb atoms. These two detection steps are performed cyclically, which effectively suppresses the error caused by detection and atomic clouds' position.

The rotation compensation of Raman laser's mirror is achieved by the 2D-PZT turntable. When π/2, π and π/2 Raman pulses reach the mirror, their wave vectors are rotated 0, $\Omega_m T$ and $2\Omega_m T$ by adjusting the tilt angle of the mirror, respectively. Because the normal direction of the mirror is orthogonal to the x′- and y′-axis of the PZT, it has a cross angle of 45° to both y-axis (south-north direction) and z-axis (vertical upward). The minor angle error between the project of the y′-axis of the PZT and the y-axis affects the measurement accuracy. The laboratory is located 16 meters underground, a laser gyroscope is used to calibrate the south-north direction with a positioning accuracy of 2°, the deviation between the projection of y′-axis in the horizontal plane and the y-axis is within 5° (limited by the error of PZT installation and positioning).

To quantitatively calibrate the error caused by the mirror, rotation compensation is applied for the mirror around x-axis and y-axis. First, the rotation around y-axis is compensated to obtain the optimal compensation angle, then the rotation around x-axis is compensated. The compensation angle in x-axis is very small, but it causes excessive compensation in y-axis. Finally, the total compensated error is evaluated by considering the error in x- and y-axis, and the theoretical estimation of overcompensation in y-axis.

### 3.2. Experimental results and discussion

After 2D rotation compensation is performed on Raman laser's mirror, the fringe contrast is significantly improved. The experimental results of fringe contrast versus mirror's rotation rate are shown in Fig. 8, where, two sets of data by exchanging Detectors A and B are included, and each set of data includes fringe contrasts of both $^{85}$Rb and $^{87}$Rb atoms. The rotation rate around y-axis varies from -2.5 $\Omega_0$ to 0.5 $\Omega_0$, and that around x-axis varies from -1.5 $\Omega_0$ to 1.5 $\Omega_0$. The best compensation value and standard deviation of rotation rate around y-axis is $(-1.036 \pm 0.025)$ $\Omega_0$, and that around x-axis is $(0.068 \pm 0.021)\Omega_0$. It should be noted that, for atomic clouds with initial temperature of 3 μK, only atoms with a relatively lower horizontal-velocity were detected by the detectors, and the actual measurements correspond to atomic temperature of 100 nK level.

The experimental results of $\Delta\eta$ versus the rotation rate of Raman laser's mirror are shown in Fig. 9, where the rotation compensation coefficients around y-axis and x-axis are $-1.1\times10^{-9}/\Omega_0$, $-1.6\times10^{-9}/\Omega_0$, respectively. The coincidences of the $^{85}$Rb and $^{87}$Rb atomic clouds in south-north and west-east directions are similar. However, the error caused by the detection difference in west-east direction is smaller than that of the position difference of $^{85}$Rb and $^{87}$Rb atomic clouds, while in south-north direction it is 3.8 times of that of the position difference of $^{85}$Rb and $^{87}$Rb atomic clouds. The intersection of two sets of data by exchanging Detectors A and B according to x-axis is near 0 (Fig.9 (b)), while that according to y-axis is far from 0 (Fig.9 (a)). This is because there is still a small installation error in the PZT rotation compensation system. When the mirror is rotated around x-axis, it will also cause a small rotation component in y-axis, so that the intersection of the two sets of data according to y-axis has an offset from the reference point $\Omega_m = -\Omega_0$. From the fringe contrast data compensated around the x-axis, the difference angle between y′-axis of the PZT and y-axis is calculated to be -2.3°, which is less than installation accuracy (5°). Due to the existence of this difference angle, a correction of $-1.1\times10^{-11}$ is added to the data after rotation compensation around the y-axis.

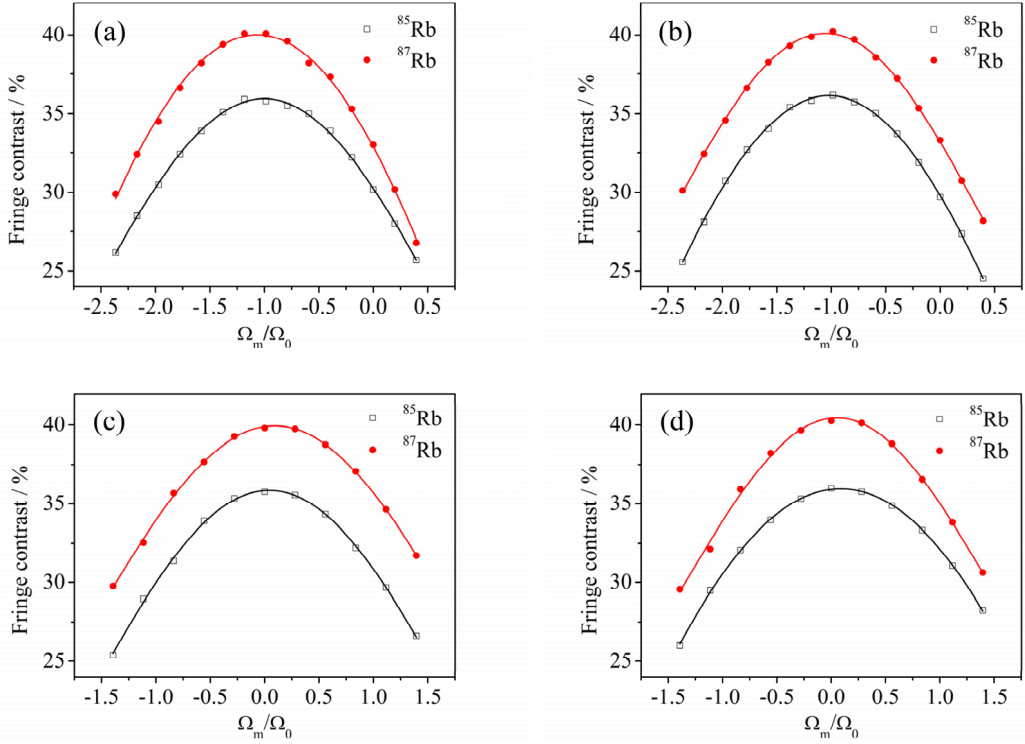

**Fig. 8** (color online) Atom interference fringes contrasts versus mirror's rotation rate. (a) Rotation compensated around y-axis, $^{85}$Rb atoms detected by Detector A and $^{87}$Rb atoms detected Detector B. (b) Rotation compensated around y-axis, $^{85}$Rb atoms detected by Detector B and $^{87}$Rb atoms detected Detector A. (c) Rotation compensated around x-axis, $^{85}$Rb atoms detected by Detector A and $^{87}$Rb atoms detected Detector B. (d) Rotation compensated around x-axis, $^{85}$Rb atoms detected by Detector B and $^{87}$Rb atoms detected Detector A.

Based on the above data, the uncertainty of rotation rate around y-axis is 0.025 $\Omega_0$ and that of Coriolis error is $1.1\times10^{-9}/\Omega_0 \times 0.025\ \Omega_0 = 2.8\times10^{-11}$; the uncertainty of rotation rate around x-axis is 0.021 $\Omega_0$ and that of Coriolis error is $1.6\times10^{-9}/\Omega_0 \times 0.021\ \Omega_0 = 3.4\times10^{-11}$. The total uncertainty of Eötvös coefficient is $\delta\eta = 4.4 \times 10^{-11}$.

## 4. Conclusion and perspectives

In summary, the Coriolis error in AI-based WEP test was analyzed theoretically and measured experimentally. An error model suitable for Coriolis effect in the 4WDR dual species AI was built, the contribution of key experimental parameters to the Coriolis phase shift was analyzed. The Coriolis phase shift introduced by position mismatch of atomic clouds and position difference of detectors was suppressed by the ADM, and the Coriolis error caused by the inconsistent initial velocity of the dual-species atomic clouds was corrected by compensating the rotation of Raman laser's mirror, and the uncertainty of $\eta$ measurement in $^{85}$Rb-$^{87}$Rb dual-species AI-based WEP test was improved to $\delta\eta = 4.4 \times 10^{-11}$. The experimentally measured error terms are consistent with the theoretical simulations. In addition, the dependence of fringe contrast on 2D rotation compensation was simulated under the condition of free evolution time of 1 s, which will be of significance for long-baseline AI-based higher-precision WEP tests in future.

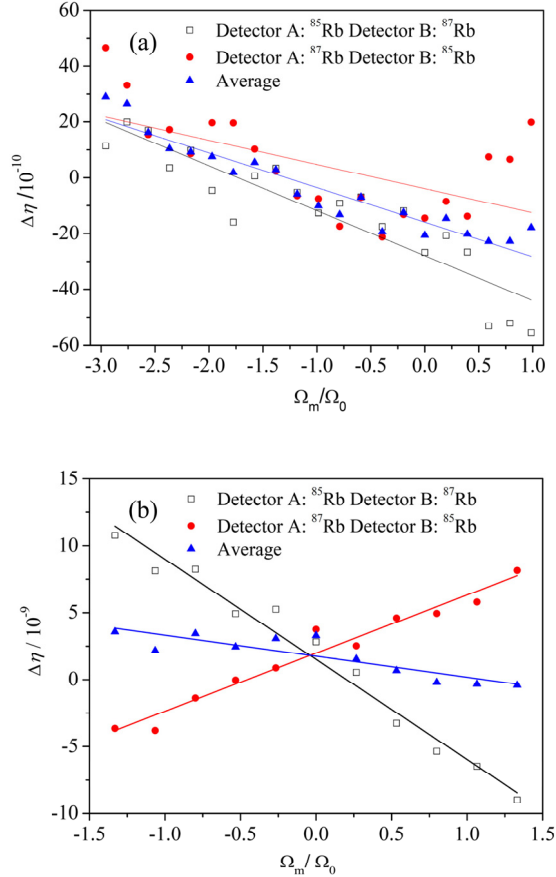

**Fig. 9** (color online) Error of $\eta$ value versus rotation rate of Raman laser's mirror. (a) Rotation compensation around y-axis. (b) Rotation compensation around x-axis. When compensating for rotation around x-axis, the compensation value around y-axis is fixed at $\Omega_m = \Omega_0$. The coefficient of rotation compensation around y-axis is $-1.1 \times 10^{-9}/\Omega_0$ and that around x-axis is $-1.6 \times 10^{-9}/\Omega_0$.

**Acknowledgment**

We acknowledge Dong-Feng Gao for his scientific discussion in error evaluation and Zong-Yuan Xiong for his technical support.

**References**
[1] Wang J 2015 *Chin. Phys. B* **24** 053702
[2] Kasevich M A and Chu S 1991 *Phys. Rev. Lett.* **67** 181
[3] Peters A, Chung K Y and Chu S 1999 *Nature* **400** 849
[4] Zhou L, Xiong Z Y, Yang W, Tang B, Peng W C, Wang Y B, Xu P, Wang J and

Zhan M S 2011 *Chin. Phys. Lett.* **28** 013701

[5] Merlet S, Bodart Q, Malossi N, Landragin A, Pereira Dos Santos F, Gitlein O and Timmen L 2010 *Metrologia* **47** L9

[6] Poli N, Wang F Y, Tarallo M G, Alberti A, Prevedelli M and Tino G M 2011 *Phys. Rev. Lett.* **106** 038501

[7] Hu Z K, Sun B L, Duan X C, Zhou M K, Chen L L, Zhan S, Zhang Q Z and Luo J 2013 *Phys. Rev. A* **88** 043610

[8] Bidel Y, Carraz O, Charrière R, Cadoret M, Zahzam N and Bresson A 2013 *Appl. Phys. Lett.* **102** 144107

[9] Wu B, Wang Z, Cheng B, Wang Q, Xu A and Lin Q 2014 *Metrologia* **51** 452

[10] Wang S K, Zhao Y, Zhuang W, Li T C, Wu S Q, Feng J Y and Li C J 2018 *Metrologia* **55** 360

[11] Huang P W, Tang B, Chen X, Zhong J Q, Xiong Z Y, Zhou L, Wang J and Zhan M S 2019 *Metrologia* **56** 045012

[12] Snadden M J, McGuirk J M, Bouyer P, Haritos K G and Kasevich M A 1998 *Phys. Rev. Lett.* **81** 971

[13] Duan X C, Zhou M K, Mao D K, Yao H B, Deng X B, Luo J and Hu Z K 2014 *Phys. Rev. A* **90** 023617

[14] Wang Y P, Zhong J Q, Song H W, Zhu L, Li Y M, Chen X, Li R B, Wang J and Zhan M S, 2017 *Phys. Rev. A* **95** 053612

[15] Canuel B, Leduc F, Holleville D, Gauguet A, Fils J, Virdis A, Clairon A, Dimarcq N, Borde C J, Landragin A and Bouyer P 2006 *Phys. Rev. Lett.* **97** 010402

[16] Tackmann G, Berg P, Schubert C, Abend S, Gilowski M, Ertmer W and Rasel E M 2012 *New J. Phys.* **14** 015002

[17] Dickerson S M, Hogan J M, Sugarbaker A, Johnson D M S and Kasevich M A 2013 *Phys. Rev. Lett.* **111** 083001

[18] Yao Z W, Lu S B, Li R B, Wang K, Cao L, Wang J and Zhan M S 2016 *Chin. Phys. Lett.* **33** 083701

[19] Yao Z W, Lu S B, Li R B, Luo J, Wang J and Zhan M S 2018 *Phys. Rev. A* **97** 013620

[20] Fray S, Diez C A, Hänsch T W and Weitz M 2004 *Phys. Rev. Lett.* **93** 240404

[21] Bonnin A, Zahzam N, Bidel Y and Bresson A 2013 *Phys. Rev. A* **88** 043615

[22] Schlippert D, Hartwig J, Albers H, Richardson L L, Schubert C, Roura A, Schleich W P, Ertmer W and Rasel E M 2014 *Phys. Rev. Lett.* **112** 203002

[23] Tarallo M G, Mazzoni T, Poli N, Sutyrin D V, Zhang X and Tino G M 2014 *Phys. Rev. Lett.* **113** 023005

[24] Zhou L, Long S T, Tang B, Chen X, Gao F, Peng W C, Duan W T, Zhong J Q, Xiong Z Y, Wang J, Zhang Y Z and Zhan M S 2015 *Phys. Rev. Lett.* **115** 013004

[25] Barrett B, Antoni-Micollier L, Chichet L, Battelier B, Lévèque T, Landragin A and Bouyer P 2016 *Nat. Commun.* **7** 13786

[26] Duan X C, Zhou M K, Deng X B, Yao H B, Shao C G, Luo J and Hu Z K 2016 *Phys. Rev. Lett.* **117** 023001

[27] Rosi G, D'Amico G, Cacciapuoti L, Sorrentino F, Prevedelli M, Zych M, Brukner Č and Tino G M 2017 *Nat. Commun.* **8** 15529

[28] Zhang K, Zhou M K, Cheng Y, Chen L L, Luo Q, Xu W J, Cao L S, Duan X C and Hu Z K 2018 arXiv:1805.07758

[29] Zhou L, He C, Yan S T, Chen X, Duan W T, Xu R D, Zhou C, Ji Y H, Bathwal S, Wang Q, Hou Z, Xiong Z Y, Gao D F, Zhang Y Z, Ni W -T, Wang J and Zhan M S 2019 arXiv:1904.07096

[30] Freier C, Hauth M, Schkolnik V, Leykauf B, Schilling M, Wziontek H, Scherneck H-G, Müller J and Peters A 2016 *J Phys.: Conf. Ser.* **723** 012050

[31] Fu Z J, Wu B, Cheng B, Zhou Y, Weng K, Zhu D, Wang Z and Lin Q 2019 *Metrologia* **56** 025001

[32] Wu X, Pagel Z, Malek B S, Nguyen T H, Zi F, Scheirer D S and Müller H 2019 *Sci. Adv.* **5** eaax0800

[33] Peters A, Chung K Y and Chu S 2001 *Metrologia* **38** 25

[34] Louchet-Chauvet A, Farah T, Bodart Q, Clairon A, Landragin A, Merlet S and Pereira Dos Santos F 2011 *New J. Phys.* **13** 065025

[35] Lan S Y, Kuan P C, Estey B, Haslinger P and Müller H 2012 *Phys. Rev. Lett.* **108** 090402

[36] Bongs K, Launay R and Kasevich M A 2006 *Appl. Phys. B: Lasers Opt.* **84** 599

[37] Farah T, Gillot P, Cheng B, Landragin A, Merlet S and Pereira Dos Santos F 2014 *Phys. Rev. A* **90** 023606

[38] Hogan J M, Johnson D M S and Kasevich M A arXiv 0806.3261v1

[39] Zhou L, Xiong Z Y, Yang W, Tang B, Peng W C, Hao K, Li R B, Liu M, Wang J and Zhan M S 2011 *Gen. Relat. Gravit.* **43** 1931

[40] Foster G T, Fixler J B, McGuirk J M and Kasevich M A 2002 *Opt. Lett.* **27** 951